\documentclass[10pt,letterpaper]{article}
\usepackage{opex3, ams math, cite}
\begin{document}

\def\ptl{ IEEE Photon.\ Technol.\ Lett.\ }
\def\spie{ Proc.\ SPIE }

\title{Study on differences between high contrast grating reflectors for TM and TE polarizations and their impact on VCSEL designs}
\author{Il-Sug Chung*}
\address{Department of Photonics Engineering (DTU Fotonik), Technical University of Denmark \\ {\O}rsteds Plads, DK-2800 Kgs. Lyngby, Denmark}
\email{*ilch@fotonik.dtu.dk}

\begin{abstract} 
A theoretical study of differences in broadband high-index-contrast grating (HCG) reflectors for TM and TE polarizations is presented, covering various grating parameters and properties of HCGs. It is shown that the HCG reflectors for TM polarization (TM HCG reflectors) have much thicker grating thicknesses and smaller grating periods than the TE HCG reflectors. This difference is found to originate from the different boundary conditions met for the electric field of each polarization. Due to this difference, the TM HCG reflectors have much shorter evanescent extension of HCG modes into low-refractive-index media surrounding the HCG. This enables to achieve a very short effective cavity length for VCSELs, which is essential for ultrahigh speed VCSELs and MEMS-tunable VCSELs. The obtained understandings on polarization dependences will be able to serve as important design guidelines for various HCG-based devices.
\end{abstract}


\section{Introduction}
The grating structure referred to as high contrast grating (HCG) is near subwavelength gratings formed in a thin high-refractive-index layer surrounded by lower refractive index materials\cite{Mateus_2004}. HCGs can provide high reflectivity to a surface-normal incident wave over a broad wavelength range, e.g. higher than 99.9\% over broader than 100 nm. This capability makes HCGs an attractive alternative to conventional distributed Bragg reflectors (DBRs) for vertical-cavity surface-emitting lasers (VCSELs). Since the first demonstration of VCSELs employing a HCG reflector \cite{Huang_2007, Boutami_2007}, many novel HCG-based VCSEL structures have been reported, featured by strong single-transverse-mode operation\cite{Chung_2008, Chung_2009, Sano_2012}, efficient MEMS tuning of emission wavelength\cite{Huang_2007b, Chung_2010}, integration onto a silicon-on-insulator (SOI) wafer\cite{Chung_2010b, Sciancalepore_2012, Park_2015}, and control of transverse mode confinement\cite{Sciancalepore_2011, Gebski_2014}. 

These HCG reflectors can be classified into TM HCG and TE HCG reflectors, depending on the polarization of an incident wave to which a HCG is highly reflective. Since both types can provide broadband high reflectivity and their reflection processes are  explained by the same theories\cite{Magnusson_2008, Karagodsky_2010}, less attention has been paid to the differences between the two types. I have found  from various TM and TE HCG cases \cite{Chung_2008b} that optimal HCG parameters such as grating thickness, period, and duty cycle, and the evanescent tail length of HCG modes into the low index media are quite different for each type. We need to note that this evanescent tail length can be a key parameter in the designs of HCG-based high-speed VCSELs and MEMS-tunable VCSELs, significantly influencing their performances as discussed in Section 4. This rather empirical observation and the applicational importance of polarization dependent properties have motivated the studies of this paper, asking two questions: whether the empirically observed polarization dependences occur by chance or result from a physical origin and whether other properties of HCG reflectors such as penetration depth, reflection phase shift, and stopband width are also polarization dependent.     

This paper is organized as follows: Firstly, the polarization dependences of HCG parameters as well as HCG reflector properties are statistically investigated in Section 2. Then, the origin of polarization dependences are investigated in Section 3. In Section 4, the importance of polarization dependent properties in designing HCG-based devices, mostly VCSELs is discussed.

For numerical investigations, the rigorous coupled wave analysis (RCWA) method was employed\cite{Moharam_1995}. I chose the design wavelength, grating material, and surrounding materials to be 1550 nm, Si with a refractive index of 3.47, and air, respectively. However, the results in this paper should be valid for other wavelengths and other sets of materials with a high refractive-index contrast. 

\begin{figure}[t!]
\centering
\includegraphics[angle=0, width=\columnwidth]{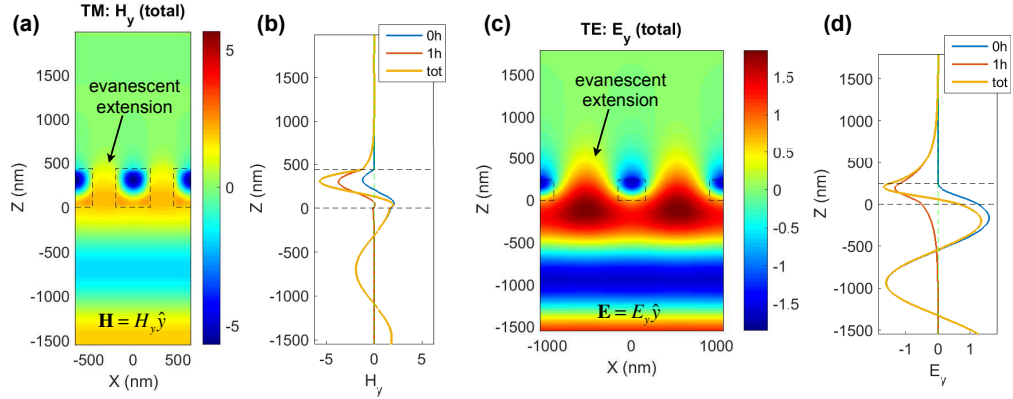}%
\caption{\label{comparison} Field profiles around (a, b) TM and (c, d) TE HCG reflectors with 1550-nm-wavelength TM and TE waves being incident from the bottom, respectively. In (b, d), the field profiles at $x$=0 nm (tot) are shown together with their 0th (0h) and 1st harmonic (1h) components. The TM HCG has a grating thickness of 420 nm, a grating period of 642 nm, and a grating bar width of 410 nm while the TE HCG has a thickness of 245 nm, a period of 1084 nm, and a width of 322 nm. The black dashed lines denote grating boundaries. The refractive index of the grating made of Si is 3.47 and both incident and exit media are air.}%
\end{figure}

\section{Statistical investigation}

\subsection{Insight from examples}
Figure 1 compares exemplary TM and TE HCG reflectors, both optimized to have a broad stopband around a wavelength of 1550 nm. In Figs. 1(a) and 1(c), the TM HCG reflector has a thicker thickness, a shorter period, and a larger duty cycle (ratio of grating bar width over grating period) than the TE HCG. Regarding the evanescent tail, the TE HCG reflector has a much longer field extension into the transmitted side (above the HCG) than the TM HCG reflector, as shown in Figs. 1(a) and 1(c). Figures 1(b) and 1(d) decomposing fields into harmonics show that the first order harmonic component 1h mainly determines the evanescent tail length and it  extends not only into the transmitted side but also into the reflected side.

\begin{figure}[t]
\centering
\includegraphics[angle=0, width=0.99\columnwidth]{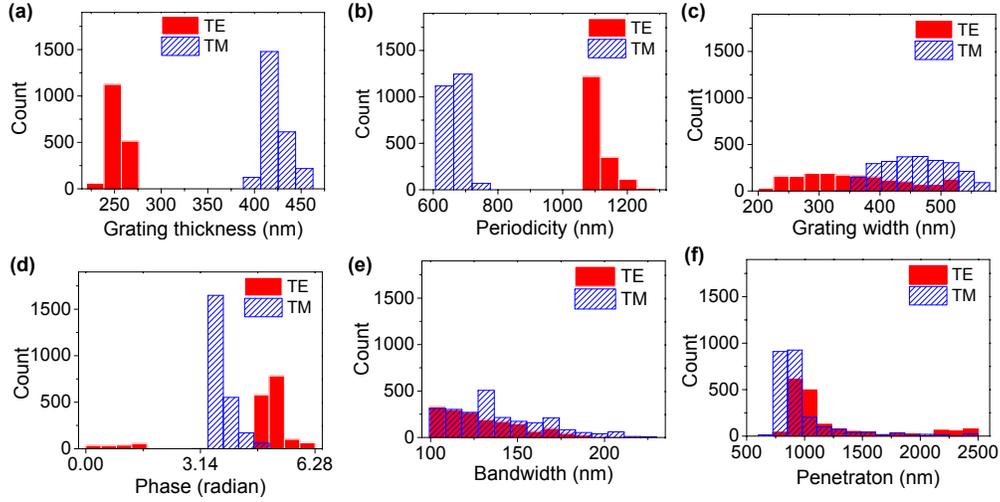}%
\caption{\label{stats} (a)-(c) Histograms of HCG parameters (grating thickness, grating period, and grating bar width). (d)-(f) Histograms of HCG properties (reflection phase, stopband width, and penetration depth).}%
\end{figure}

\begin{figure}[t]
\centering
\includegraphics[angle=0, width=0.99\columnwidth]{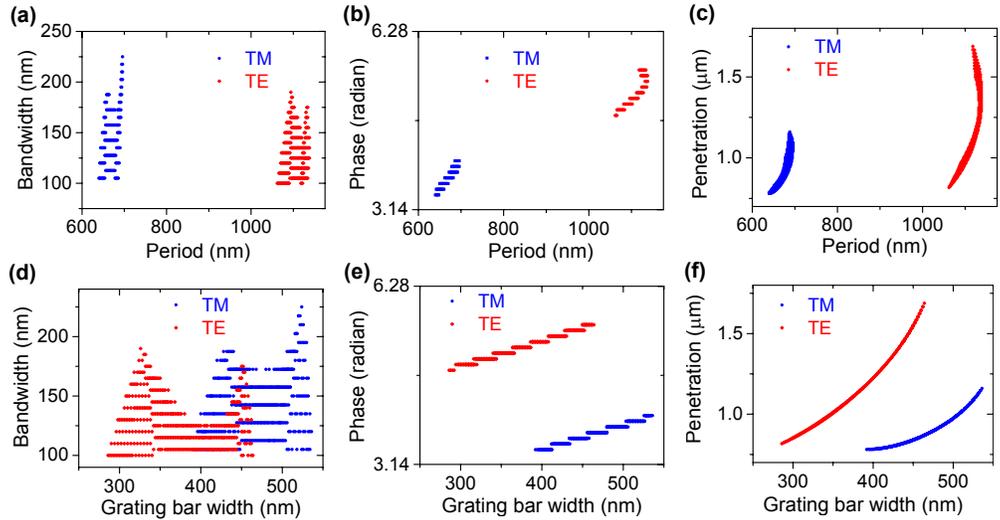}%
\caption{\label{correlation} Plots of HCG properties (stopband width, reflection phase, and penetration) as a function of HCG parameters (grating period and grating width) at a specific grating thickness. The thicknesses of TE HCGs and TM HCGs are 245 nm and 420 nm, respectively. }%
\end{figure}

\subsection{Statistical analysis}
In order to find out whether the polarization dependences observed for the exemplary HCG reflectors are universal, a statistical investigation is conducted. A broad range of TM and TE HCG structures are simulated by varying the HCG parameters, i.e., grating thickness, grating period, and grating bar width independently over very broad ranges. Among the simulated parameter sets, only those HCG parameter sets that result in a reflectivity higher than 99.9 \% throughout a stopband broader than 100 nm around 1550-nm wavelength are counted for analysis. The histograms in Fig. 2 summarize the simulation results:
\begin{itemize}
\item Figures 2(a) to 2(c) show the polarization dependence of HCG parameters. As in the exemplary HCGs, the TE HCGs have thinner thicknesses and longer periods than the TM HCGs. The grating thickness and period values are concentrated around specific values while the the grating bar width values are widely distributed. Thus, one may guess that there could be conditions to be met for grating thickness and period, which are investigated in Section 3. 
\item Figures 2(d) to 2(f) show the polarization dependence of HCG properties, i.e., reflection phase, stopband width, and penetration length\cite{Chung_2010, Zhao_2010}. The TM HCGs are more likely to have a smaller reflection phase and a smaller penetration. Especially, the smaller penetration leads to a shorter effective cavity length advantageous for high-speed VCSELs and MEMS-tunable VCSELs, as discussed in Section 4. Regarding the stopband width, TM and TE HCGs have similar distributions.  
\end{itemize}

To find out whether there is an apparent relationship between the HCG properties and HCG parameters, the HCG properties are plotted in Fig. 3 as a function of grating period and grating bar width at a given grating thickness:
\begin{itemize}
\item As shown in Fig. 3(f), the penetration depth monotonically increases with the grating bar width. It can be explained in this way. As the grating bar width increases, the coupling between neighboring grating bars increases. Since the penetration depth is related to the lateral propagation distance within a HCG layer, the increased coupling may lead to a longer penetration length. The reflection phase also gradually increases with the grating bar width as shown in Fig. 3(e).
\item The bandwidth is hardly related to either the grating period nor bar width, as shown in Figs. 3(a) and 3(d).  
\item Other correlations shown in Figs. 3(b) and 3(c) are not clear and sometimes change at different grating thicknesses.
\end{itemize}

\begin{figure}[t!]
\begin{center}
\includegraphics[angle=90, width=0.62\columnwidth]{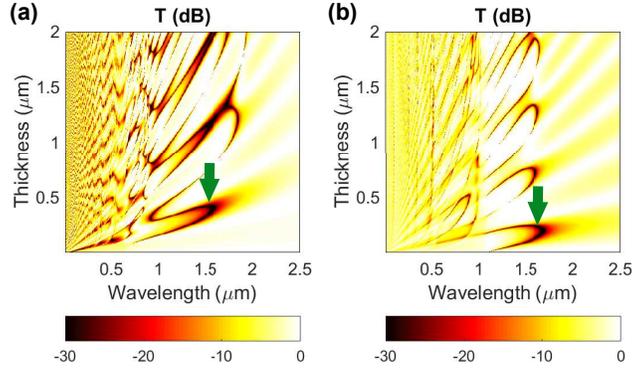}%
\caption{\label{dispersions} Transmittance plots of (a) TM HCGs and (b) TE HCGs as a function of wavelength and grating thickness. The grating period and width are the same as in Fig. 1.}%
\end{center}
\end{figure}

\section{Origin of polarization dependence}

In this section, the origin of polarization dependences observed in grating thickness, grating period, and evanescent tail length are  investigated.

\subsection{Grating thickness}
In Fig. 4, the transmittances of the exemplary TM and TE HCGs of Fig. 1 are plotted as a function of wavelength and grating thickness. Broad bandwidths can be obtained at the bendings of low transmittance line \cite{Karagodsky_2010}. For both TM and TE HCGs, the convex bending occurring at the smallest thickness, designated by a green arrow gives the flattest curvature corresponding to the broadest bandwidth: other bendings with a thicker thickness do not provide a stopband broader than 100 nm. This explains why there is a single (not multiple) optimal grating thickness for each polarization in Fig. 2(a). The reason this bending occurs at a shorter thickness in the TE HCG than the TM HCG can be explained by investigating the eigenmode profiles.  

\begin{figure}
\centering
\includegraphics[angle=-90, width=1.05\columnwidth]{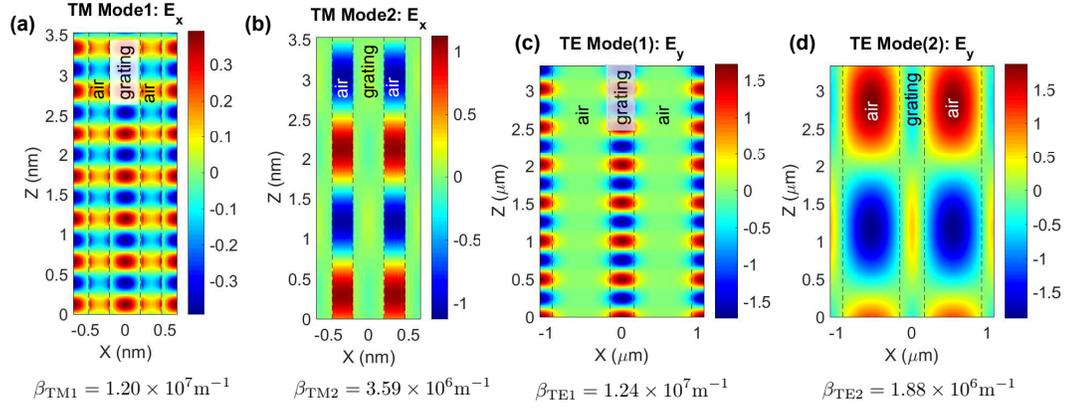}%
\caption{\label{mode_profiles} Eigenmode electric-field profiles $E_{t,m}$ of (a, b) the TM waveguide array and (c, d) the TE waveguide array, with propagation constant, $\beta$ values. The waveguide period and width are the same as given in the caption of  Fig. 1. The black dashed lines designate waveguide boundaries.}%
\end{figure}

Figure 5 shows the field profiles of the two propagating eigenmodes of the TM and TE waveguide arrays (along $y$ axis) at a wavelength of 1550 nm, which are infinite versions (along $z$ direction) of the TM and TE HCGs. There are only two propagating modes for both TM and TE cases as the considered wavelength is within the so-called two-mode regime\cite{Karagodsky_2010}. To have a transmittance close to 0, the lateral average of these two modes at the grating exit plane should cancel each other. This condition can be expressed by\cite{Karagodsky_2010}, 
\begin{equation}\label{eq1}
t_0 \propto \sum_{m=1,2} (a_m + a_m^\rho)\Lambda^{-1} \int_0^\Lambda E_{t,m}(x)e^{-i\beta_m z_\text{ex}}dx = 0,
\end{equation} 
where $t_0$ is the transmittance amplitude of the 0-th harmonic component, $a_m$ and $a_m^\rho$ are the coefficients of the upward- and downward-moving $m$-th modes at the exit plane, respectively, $\Lambda$ is grating period, $E_{t,m}$ is the normalized electric field component ($E_{x,m}$ for TM and $E_{y,m}$ for TE) of the $m$-th mode, and $z_\text{ex}$ is the $z$ coordinate of the exit plane. This $z_\text{ex}$ value becomes same as the grating thickness, $t_\text{gr}$ if the input plane of the grating is assumed to be at $z$=0. Since the two propagating modes have similar amplitudes, i.e., $a_1 \sim a_2$ and $a_1^\rho \sim a_2^\rho$, Eq. (\ref{eq1}) can be simplified: 
\begin{equation}\label{eq2}
\begin{split}
t_0 & \propto (a_1+a_1^\rho)\Lambda^{-1} \sum_{m=1,2} e^{-i\beta_m t_\text{gr}}\int_0^\Lambda E_{t,m}(x)dx \\
       & = (a_1+a_1^\rho)\Lambda^{-1} \sum_{m=1,2} e^{-i\beta_m t_\text{gr}}\times 1\\
       & = (a_1+a_1^\rho)\Lambda^{-1} e^{-i\beta_2 t_\text{gr}}\left(e^{-i(\beta_1-\beta_2) t_\text{gr}}+1\right),      
\end{split}
\end{equation} 
where the normalization condition is used in the derivation from the first line to the second.  In order to make $t_0$ close to 0 in the last line of Eq. (\ref{eq2}), $(\beta_1-\beta_2)t_\text{gr}$ should be an odd integer multiple of $\pi$. Thus, the thinnest thickness, $t_\text{gr,0}$ is given by 
\begin{equation}\label{eq3}
t_\text{gr,0} = \frac{\pi}{\beta_1-\beta_2}.
\end{equation}

Substituting $\beta_m$ values of the TE and TM waveguides given in Fig. 5 into Eq. (\ref{eq3}) gives $t_\text{gr,0}$ values of 299 nm and 374 nm for the TE and TM HCGs, respectively. These values are similar as the exact values given in the caption of Fig. 1. The difference in the optimal thicknesses of TM and TE HCG reflectors is attributed to their difference in $(\beta_1-\beta_2)$. 

This difference in $(\beta_1-\beta_2)$ originates from the polarization-dependant boundary conditions. In Fig. 5, both TM mode 1 and TE mode 1 have more electric field confined within the grating bars than the air, while both TM and TE modes 2 have more in the air. In the TM case, the continuity of $D_x$(=$\epsilon E_x$) is required at the interface of grating and air. As shown in Fig. 5(a), this results in a large enhancement of the electric field in the air region for the TM mode 1 case, due to the large refractive index difference between the grating and air. However, in the TE mode 1 case, the continuity of $E_y$ is required at the grating-air interface, which results in no enhancement as shown in Fig. 5(c). Since the propagation constant, $\beta_m$ is proportional to the effective index of the mode, averaged with the electric field intensity as a weighing factor, the effective index of the TM mode 1 cannot be high: considerable electric field intensity is also distributed in the air while in other cases, the effective indices are determined solely by the filling factor. Therefore, $(\beta_1-\beta_2)$ becomes smaller for the TM case.

\begin{figure}
\begin{center}
\includegraphics[angle=-90, width=0.64\columnwidth]{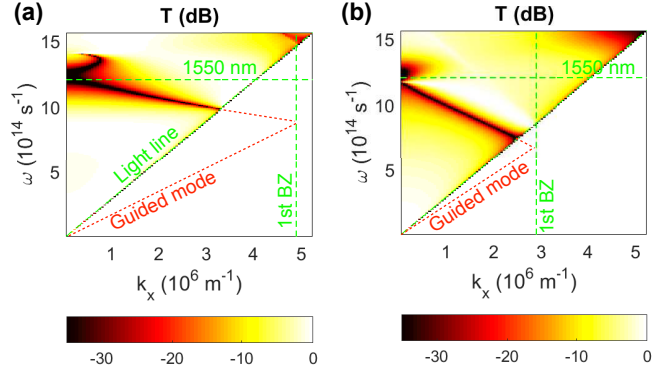}%
\caption{\label{dispersions} Transmittances of (a) the TM HCG and (b) the TE HCG as a function of wavevector, $k_x$ and angular frequency, $\omega$. The green dashed lines designate 1550-nm wavelength, light line, and the 1st Brillouin zone boundary. The red dashed lines designate the guesses for the lowest guided modes below the light line.}%
\end{center}
\end{figure}

\subsection{Grating period}

HCGs can be approximated as a slab waveguide with waves propagating along $x$ direction. The transmittance plots in Fig. 6 show the dispersion of this waveguide, i.e., $\omega$ vs. $k_x$ above the light line\cite{Yang_2013}. Due to the periodicity, the waveguide dispersion is folded at the first Brillouin zone boundary, i.e., $k_x$=$\pi/\Lambda$. Since $k_z\sim\pi/t_\text{gr}$ for the lowest waveguide mode (resonance condition along $z$ direction), the $k_x$ value can be approximately evaluated in the unfolded dispersion curve by:
\begin{equation}\label{eq4}
\begin{split}
\left(\frac{\omega}{c} \right)^2 &= k_x^2 + k_z^2 \\
&\sim k_x^2 + \left(\frac{\pi}{t_\text{gr}}\right)^2.
\end{split}
\end{equation}
In order to have a crossing with the $\omega$ axis at ($k_x$, $\omega$)=(0, $2\pi/1550$ nm) in the folded dispersion curve as shown in Fig. 6, Eq.  (\ref{eq4}) should be satisfied at ($k_x$, $\omega$)=($2\pi/\Lambda$, $2\pi/1550\text{ nm}$):

\begin{equation}\label{eq5}
\end{equation}
Equation (\ref{eq5}) shows that the larger period of the TE HCG can be attributed to its thinner thickness. Since the thinner thickness of the TE HCG originates from the boundary condition, one may say that the polarization dependent boundary condition leads to the polarization dependent differences in period as well as thickness observed in Figs. 2(a) and 2(b).

\subsection{Evanescent tail}
The grating period determines this evanescent tail length. The propagation constant of the $m$-th order harmonic component in the air, $k_z^\mathrm{air}$ is given by, 
\begin{equation}\label{eq:evanescent}
k_z^\mathrm{air}=\sqrt{\left(2\pi/\lambda\right)^2-\left(2\pi m/\Lambda\right)^2},
\end{equation}
where $\lambda$ is the wavelength of an incident wave and $\Lambda$ is the grating period. Since $\Lambda<\lambda$, all higher order components than the zeroth decay evanescently. Equation (\ref{eq:evanescent}) tells us that a smaller $\Lambda$ gives a larger purely imaginary $k_z^\text{air}$, leading to a shorter evanescent tail.  This explains why the TE HCG with a larger $\Lambda$ has a longer evanescent tail. 

\begin{figure}[t!]
\centering
\includegraphics[angle=0, width=0.62\columnwidth]{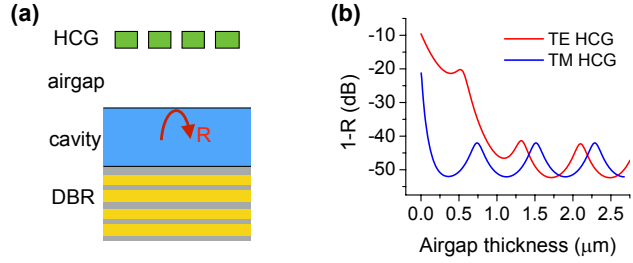}%
\caption{\label{comparison} Transmittance, $(1-R)$ of the top reflector seen from the cavity layer as a function of the airgap thickness. A TE- and a TM-polarized plane wave at a wavelength of 1550 nm are incident from the cavity layer for the TE HCG and the TM HCG cases, respectively. The same grating parameters as in Fig. 1 are assumed. The refractive index of the cavity layer made of InP is 3.1661. }%
\end{figure}

\section{Discussion for HCG-based VCSEL designs}
As shown in Fig. 7(a), HCG reflectors require a low-refractive-index gap layer between a HCG layer and a cavity layer when incorporated into VCSELs. The reflectivity, $R$ seen from the cavity layer is a composite reflectivity including the reflections from the HCG as well as the air gap/cavity interface. If the air gap thickness is close to the evanescent tail length, the HCG mode field may couple to the cavity layer, which reduces $R$. Figure 7(b) compares the $R$ values of the TM and TE HCGs of Fig. 1 as a function of the air gap thickness. The TM HCG reaches the saturated reflectivity value at an air gap thickness of 0.3 $\mu$m while the TE HCG does at 1.7 $\mu$m. This is consistent with the observation that the TM HCG has a shorter penetration tail than the TE HCG. Note that the air gap thickness of 0.3 $\mu$m is 0.19 times of the wavelength, which is even thinner than the optical thickness of a single DBR layer. 

This thin gap thickness achievable for TM HCG allows TM HCG-based VCSELs to achieve a larger confinement factor and a shorter effective cavity length than conventional DBR-based VCSELs. The larger confinement factor reduces the threshold current, $I_\mathrm{th}$, increasing the modulation speed through the dependence on stimulated emission current, $(I-I_\mathrm{th})^{1/2}$. This increase is more significant for low-enegy-consuming VCSELs for chip-level optical interconnects where the operating current, $I$ is close to $I_\mathrm{th}$\cite{Hofmann_2012}. Furthermore, the shorter effective cavity length is a key factor for achieving a higher modulation speed through a shorter photon lifetime as well as for increasing the tuning range of HCG-based MEMS tunable VCSELs\cite{Chung_2010}. For example,  a TM HCG-based VCSEL can achieve a modulation speed higher than 100 Gb/s by using the findings of this paper \cite{Park_2015}.

The shorter period of TM HCGs is more adequate for wavefront engineering, such as beam focusing and tilting since it can make the phase front smoother. The reason is that the spatial variation of the reflection or transmission phase from a HCG is obtained by discrete phase variation from each period.  

As shown in Fig. 2(f), TE HCGs can be designed to have a long penetration depth, which is advantageous for achieving a large lateral coupling efficiency from the vertical cavity to an in-plane waveguide\cite{Chung_2010b}. The shorter penetration length of TM HCGs can allows for a short effective cavity length of HCG-based MEMS-tunable VCSELs, leading to a large tuning range\cite{Chung_2010}.  

\section{Conclusion}
In conclusion, all of HCG parameters and some of HCG properties are different for the TM HCG and TE HCG broadband reflectors. Among HCG parameters, the grating thickness and grating period are quite different and these differences are found to originate from different electric field boundary conditions met for each type of HCG reflector. Among HCG properties, the reflectivity and stopband width are almost same for both types while the evanescent tail length and reflection phase are significantly different and the penetration depth is slightly different for each type. Furthermore, it is found that the grating period and the grating bar width determine the evanescent tail length and the penetration depth, respectively. It should be noted that the polarization dependent HCG properties significantly influence the important properties of HCG-based VCSEL cavities including effective cavity length and quantum confinement factor. Thus, they need to be considered in designing ultrahigh-speed VCSELs and MEMS-tunable VCSELs.

\section*{Acknowledgements}
The author gratefully acknowledges support from the Danish Research Council through the FTP project (Grant No. 0602-01885B). 

\end{document}